\documentclass[12pt]{article}
\newcommand {\be}{\begin{equation}}
\newcommand {\ee}{\end{equation}}
\usepackage{graphicx}

\begin{document}

\title{Hamiltonian Quantum Dynamics With Separability Constraints }
 \author{Nikola Buri\' c  \thanks{e-mail: buric@phy.bg.ac.yu}
\\
Institute of Physics,
 P.O. Box 57, 11001 Belgrade, Serbia.}

 \maketitle

\begin{abstract}
 Schroedinger equation on a Hilbert space ${\cal
H}$, represents a linear Hamiltonian dynamical system on the space
of quantum pure states, the projective Hilbert space $P {\cal H}$.
Separable states of a bipartite quantum system form a special
submanifold of $P {\cal H}$.
 We analyze the Hamiltonian dynamics  that
 corresponds to the quantum system constrained on the manifold of
 separable states, using as an important example the
 system of two interacting qubits. The constraints introduce nonlinearities which
 render the dynamics nontrivial. We show that the
 qualitative properties of the constrained dynamics clearly
 manifest the symmetry of the qubits system.
  In particular, if the quantum Hamilton's operator has not enough symmetry,
   the constrained dynamics is nonintegrable, and displays the
   typical features of a Hamiltonian dynamical system with mixed phase
   space. Possible physical realizations of the separability
   constraints are discussed.

\end{abstract}


\newpage

\section{Introduction}

Classical and quantum descriptions of a physical system that is
considered as composed of interacting subsystems have radically
different features. The typical feature of quantum dynamics is the
creation of specifically quantum correlations, the entanglement,
among the subsystems. On the other hand, the typical property of
classical description is the occurrence of chaotic orbits and
fractality of the phase space portrait, which can be considered as
typically classical type of correlations between the subsystems.
The type of correlations introduced by the dynamical entanglement
does not occur in the classical description, and likewise, the
type of correlations introduced by the chaotic orbits with fractal
structures does not occur in the quantum description. This
intriguing complementarity of the two descriptions represents a
problem that is expected to be solved by a detailed formulation of
the correspondence principle.

Comparison of typical features of classical and quantum mechanics
is facilitated if the same mathematical framework is used in both
theories. It is well known, since the work of Kibble
\cite{Kibble1},\cite{Kibble2},\cite{Kibble3}, that the quantum
evolution, determined by the linear Schroedinger equation, can be
represented using the typical language of classical mechanics,
that is as a Hamiltonian dynamical system on an appropriate phase
space, given by the Hilbert space geometry of the quantum system.
This line of research was later developed into the full geometric
Hamiltonian representation of  quantum mechanics.
\cite{Heslot}-$[12]$. Such geometric formulation of quantum
mechanics has recently inspired natural definitions of measures of
the entanglement \cite{Hughston4}, and has been used to model  the
spontaneous collapse of the state vector
\cite{Hughston},\cite{Adler}, and dynamics of decoherence
\cite{Izrael}.

It is our goal to use the geometric Hamiltonian formulation of
quantum mechanics to study the relation between the dynamical
entanglement and typical qualitative properties of Hamiltonian
dynamics.  Motivated by the fact that the Schroedinger equation
can always be considered as a Hamiltonian dynamical system, and
that for Hamiltonian systems the definitions and properties of the
dynamical chaos are well understood, we shall seek for a formal
condition that when imposed on the Hamiltonian system representing
the Schroedinger equation of the compound quantum system renders
the Hamiltonian dynamics nonintegrable and chaotic. It is well
known that the linear Schroedinger equation of quantum mechanics
represents always an integrable Hamiltonian dynamical system,
irrespective of the dynamical symmetries of the system.
 This is in
sharp contrast with the Hamiltonian formulation of classical
systems, where enough symmetry implies integrability and the lack
of it implies the chaotic dynamics. Linearity of the quantum
Hamiltonian dynamics, and the consequent integrability, is
introduced in the Hamiltonian formulation by a very large
dimensionality of the phase space of the quantum system. This high
dimensionality can be considered as a consequence of two reasons.
For a single quantum system, say a one dimensional particle in a
potential, linear evolution and with it the principle of state
superposition require infinite dimensional phase space of the
Hamiltonian formulation. If the classical mechanical model is
linear, say the harmonic oscillator, the quantum Hamiltonian
dynamics can be exactly describe on the reduced finite-dimensional
phase space, the real plane in the case of the harmonic
oscillator. The other related reason that increases the
dimensionality of the quantum phase space compared to the
classical model is in the way the state space of the compound
systems are formed out of the components state spaces in the two
theories. In order to represents the entangled states as points of
the quantum phase space the dimensionality of the quantum phase
space is much larger than just the sum of the dimensions of the
components phase spaces. The points in the Cartesian product of
the components phase spaces represent the separable quantum states
and form a subset of the full quantum phase space. Needles to say,
 although the separable states are the most classical-like states of the
compound system, they still are quantum states with nonclassical
properties like nonzero dispersion of some subsystem's variables.
Our main result will be that when the quantum dynamics,
represented as a Hamiltonian system, is constrained on the
manifold of separable quantum states the relation between the
symmetry and the qualitative properties of the dynamics such as
integrability or chaotic motion is reestablished. Thus,
suppression of
 dynamical entanglement is enough to enable manifestations of the qualitative differences in dynamics of
 quantum systems and the relation between integrability and symmetry,
  traditionally related with classical mechanical models.

In order to study the relation between the dynamical entanglement,
separability and the properties of Hamiltonian formulation of the
quantum dynamics we shall use, in this paper, the simplest quantum
system that displays the dynamical entanglement, that is a system
of two interacting qubits:
\begin{equation}
  H= {\omega} { \sigma}^1+ {\omega}{\sigma}^2+ \mu_x{{
\sigma_x}}^1{ \sigma_x}^2+\mu_y{{ \sigma_y}}^1{
\sigma_y}^2+\mu_z{{ \sigma_z}}^1{ \sigma_z}^2,
\end{equation}
 where  ${\sigma_{x,y,z}}^i$ are the three Pauli matrices  of the
$i$-th qubit, and satisfy the usual $SU(2)$ commutation relations.
In particular we shall compare the dynamics of the system (1) in
the case $\mu_z\neq 0, \mu_x=\mu_y=0$ with the case when
$\mu_x\neq 0, \mu_z=\mu_y=0$. The former case is symmetric with
respect to $SO(2)$ rotations around $z$-axis and the later  lacks
this symmetry.
 Besides its simplicity, the systems of the form (1) are
of considerable current interest because the hamiltonian of the
universal quantum processor is of this form \cite
{Duch},\cite{QIP}.

  Various lines of research, during the last decade,
improved the understanding of the relation between dynamical
entanglement and properties of the dynamics. Strong impetus to the
study of all aspects of quantum entanglement came from the theory
of quantum computation \cite{QIP}. Quantization of classical
non-integrable systems, and various characteristic properties of
resulting quantum systems, have been studied for a long time
\cite{quantchaos1}. The dependence of the dynamical entanglement,
between a quantum system and its environment, on the qualitative
properties of the dynamics of the system  was studied indirectly,
within the theory of environmental decoherence \cite{decoh1}. The
relation between the rates of dynamical entanglement and the
qualitative properties of the dynamics in the semi-classical
regime was initiated in the reference \cite{Furuya} and  various
aspects of this relations have been studied since
\cite{Miller}-\cite{Anals}. The relation between the symmetry of
the genuinely quantum system (1) and the degree of dynamical
entanglement was studied in reference \cite{ja1}. As we shall see,
our present analyzes is related to the quoted works, but the
relation between the dynamical entanglement and symmetry is here
approached from a very different angle

The structure of the paper is as follows. We shall first
recapitulate the necessary background such as: the complex
symplectic and Riemannian geometry of ${\textbf{C}}P^n$;
Hamiltonian formulation on ${\textbf{C}}P^n$ of the quantum
dynamics; geometric formulation of the set of separable pure
states and Hamiltonian formulation of the constrained dynamics. In
parallel with the general reminder, the explicit formulas for the
system of two interacting qubits will be given.
 These are then
applied, in section 3, to the study the qualitative properties of
the separability constrained dynamics for the qubits systems. The
main results are summarized and discussed in section 4. There we
also discuss a model of an open quantum system with dynamics that
 clearly  differentiates between the symmetric and the
 nonsymmetric systems.

\section{Geometry of the state space ${\textbf C}P^n$}

Hamiltonian formulation of quantum mechanics is based on the fact
that the scalar product of vectors $|\psi>$ in the Hilbert space
of a quantum system  can be used to represent the linear
Schroedinger equation of quantum mechanics in the form of
Hamilton's equations. The canonical phase space structure of this
equations is determined by the imaginary part of the scalar
product, and the Hamilton' s function is given by the quantum
expectation $<\psi|H|\psi>$ of the quantum hamiltonian.

However, due to phase invariance and arbitrary normalization the
proper space of pure quantum states is not the Hilbert space used
to formulate the Schroedinger equation, but the projective Hilbert
space which is the manifold to be used in the Hamiltonian
formulation of quantum mechanics. In general, the resulting
Hamiltonian dynamical system is infinite-dimensional, but we shall
need the general definitions only for the case of  quantum system
with finite-dimensional Hilbert space, like the finite collection
of qubits, in which case the quantum phase space is also
finite-dimensional.
 We shall first review the
definition of the complex projective space ${\textbf C}P^n$, and
then briefly state the basic definitions and recapitulate the
formulas which are needed for the Hamiltonian formulation of the
quantum dynamics on the state space and its restriction on the
separable state subset. The general reference for the mathematical
aspects of complex differential geometry is \cite{Kobajasi}.  All
concepts and formulas will be illustrated using the system of two
interacting qubits.

Differential geometry of the state space ${\textbf C}P^n$ is
discussed by viewing it as a real $2n$ dimensional manifold
endowed with complex, Riemannian and symplectic structure. In the
case of ${\textbf C}P^n$ this three structures are compatible.

\subsection{Definition and intrinsic coordinates of ${\textbf C}P^n$
}

States of a collection of $N=n+1$ qubits are represented using
normalized vectors of the complex Hilbert space ${\textbf{C}}^N$.
Since all quantum mechanical predictions are given in terms of the
Hermitian scalar product on ${\textbf{C}}^N$, and this is
invariant under multiplication by a constant (vector independent)
phase factor, the states of the quantum system are actually
represented by equivalence classes of vectors in ${\textbf{C}}^N$.
Two  vectors $\psi_1$ and $\psi_2$ are equivalent: $\psi_2\sim
\psi_1$ if there is a complex scalar $a\neq 0$ such that
$\psi_2=a\psi_1$. This set of equivalence classes defines the
complex projective space: ${\textbf{C}}P^n:\equiv
({\textbf{C}}^{n+1}-0)/\sim$. It is the state space of the system
of $N$ qubits.
 Global
coordinates $(c^1,\dots c^N)$ of a vector in ${\textbf{C}}^N$ that
represent an equivalence class $[\psi]$, that is an element of
${\textbf{C}}P^n$, are called homogeneous coordinates on
${\textbf{C}}P^n$. The complex projective space is topologically
equivalent to $S^{2n+1}/S^1$, where the $2n+1$-dimensional sphere
comes from normalization and the circle $S^1$ takes care of the
unimportant overall phase factor.

The projective space ${\textbf{C}}P^n$ is locally homeomorphic
with ${\textbf{C}}^n$.  Intrinsic coordinates on ${\textbf{C}}P^n$
are introduced as follows. A chart $U_{\mu}$ consists of
equivalence classes  of all vectors in $({\textbf{C}}^{n+1}-0)$
such that $c_{\mu}\neq 0$. In the chart $U_{\mu}$ the local ( so
called inhomogeneous) coordinates $\zeta^{\nu},\> \nu=1,2\dots n$
are given by:
\begin{equation}
\zeta^{\nu}=\xi^{\nu} \quad (\nu\leq\mu-1), \quad
\zeta^{\nu}=\xi^{\nu+1}\quad \quad (\nu >\mu),
\end{equation}
where
 \begin{equation}
 \xi^{\nu}=c^{\nu}/c^{\mu}\quad
\nu=1,2,\dots \mu-1,\mu+1,\dots  n+1.
\end{equation}

The coordinates $\zeta_{\mu}^{\nu}(c)$ and $\zeta_{\mu'}^{\nu}(c)$
of a point $c$ which belongs to the domain where two charts
$U_{\mu}$ and $U_{\mu'}$ overlap are related by the following
holomorphic transformation
\begin{equation}
\zeta_{\mu'}^{\nu}(c)=(c^{\mu}/c^{\mu'})\zeta_{\mu}^{\nu}(c)
\end{equation}

As an illustration consider the system of two qubits. The Hilbert
space is ${\cal H}={\cal H}_1\bigotimes {\cal
H}_2={\textbf{C}}^2\otimes {\textbf{C}}^2={\textbf{C}}^4$. As a
basis we can choose the set of separable vectors
$|\uparrow\uparrow>,\>|\uparrow\downarrow>,|\downarrow\uparrow>,|\downarrow\downarrow>$
or any other four orthogonal vectors. The coordinates of a vector
in ${\textbf{C}}^4$ with respect to a basis are denoted
$(c^1,c^2,c^3,c^4)$. The corresponding projective space is
${\textbf{C}}P^3\equiv S^{7}/S^1$. At least two charts are needed
to define the intrinsic coordinates over all ${\textbf{C}}P^3$.
Consider first all vectors with a nonzero component along
$|1>=|\uparrow\uparrow>$ that is $c^1\neq 0$, i.e. all vectors
except the vector $|\downarrow \downarrow>$. Then the numbers
$\xi^{\nu}_{1}$ are defined as $\xi_{1}^1=c^1/c^1=1,
\>\xi_{1}^2=c^2/c^1,\>\xi_{1}^3=c^3/c^1,\>\xi_{1}^4=c^4/c^1$ and
finally the three intrinsic coordinates
$(\zeta_{1}^1,\zeta_{1}^2,\zeta_{1}^3)$ are given by relabelling
of $\xi^{\nu}_{1}$:
$\zeta_{1}^1=\xi_{1}^2,\>\zeta_{1}^2=\xi_{1}^3,\>\zeta_{1}^3=\xi_{1}^4$.
To coordinatize the vector $|4>=|\downarrow \downarrow>$ we need
another chart.

Quantum mean values of linear operators on ${\textbf{C}}^4$ are
indeed reduced to functions on ${\textbf{C}}P^3$. For example,
consider the following Hamiltonian operator
\begin{equation}
H=\omega\sigma_z\otimes {\textbf{1}}+\omega{\textbf{1}}\otimes
\sigma_z+\mu\sigma_x\otimes\sigma_x
\end{equation}
In the separable bases the normalized quantum expectation
 $<\psi|H|\psi>/<\psi|\psi>$ is given by the following function of
$(c^1,c^2\dots, \bar c^4)$
\begin{equation}
H={2\omega(c^1\bar c^1-c^4\bar c^4)+
 \mu(\bar c^2c^3+\bar c^3 c^2+\bar c^1 c^4+\bar
c^4 c^1)\over c^1\bar c^1+c^2\bar c^2+c^3\bar c^3+c^4\bar c^4}.
\end{equation}
In the intrinsic coordinates $\zeta^1,\zeta^2,\zeta^3$ and their
conjugates this expression is given by
\begin{equation}
H={2\omega(1-\zeta^3\bar\zeta^3)+
\mu(\bar\zeta^1\zeta^2+\bar\zeta^2\zeta^1+\zeta^3+\bar\zeta^3)
\over 1+\zeta^1\bar \zeta^1+\zeta^2\bar \zeta^2+\zeta^3\bar
\zeta^3}.
\end{equation}

We shall also analyze the following Hamiltonian
\begin{equation}
H=\omega\sigma_z\otimes {\textbf{1}}+\omega{\textbf{1}}\otimes
\sigma_z+\mu\sigma_z\otimes\sigma_z,
\end{equation}
whose normalized mean value is given by
\begin{eqnarray}
H&=&{2\omega (c^1\bar c^1-c^4\bar c^4)+\mu (c^1\bar c^1+c^4\bar
c^4-c^2\bar c^2-c^3\bar c^3)\over c^1\bar c^1+c^2\bar c^2+c^3\bar
c^3+c^4\bar c^4} .
\end{eqnarray}
 The corresponding function on
${\textbf{C}}P^3$ is, in the intrinsic coordinates, given by
\begin{eqnarray}
H={ \omega(1-\zeta^3\bar \zeta^3)+\mu (1+\zeta^3\bar
\zeta^3-\zeta^1\bar \zeta^1-\zeta^2\bar \zeta^2)\over
1+\zeta^1\bar \zeta^1+\zeta^2\bar \zeta^2+\zeta^3\bar \zeta^3}.
\end{eqnarray}

\subsubsection{ Submanifold of  separable states}

Consider two quantum systems {\bf A} and {\bf B} with the
corresponding Hilbert spaces $ {\cal H}_A$ and $ {\cal H}_B$.
Taken together, the systems {\bf A} and {\bf B} form another
quantum system. The statistics of measurements that could be
performed on this compound system requires that the Hilbert space
of the compound system is given by the direct product ${\cal
H}_{AB}= {\cal H}_A \otimes{\cal H}_B$. The space of pure states
of the compound system is the projective Hilbert space $P{\cal
H}_{AB}$. In the case of finite dimensional state spaces ${P\cal
H}^{n+1}_A=CP^n$ and ${P\cal H}^{m+1}_A=CP^m$ the state space of
the compound system is $CP^{(m+1)(n+1)-1}$.  Vectors in ${\cal
H}_{AB}$ of the form $\psi_A\otimes\psi_B$ where $\psi_{A/B}\in
{\cal H}_{A/B}$ are called separable. The corresponding separable
states form the $(m+n)$-dimensional submanifold $CP^m\times CP^n$
embedded in $CP^{(m+1)(n+1)-1}$.

In the case of two qubits the submanifold of the separable states
$CP^1\times CP^1$ forms a quadric in the full state space
$CP^{3}$, given in terms of the homogeneous coordinates
$(c^1,c^2,c^3,c^4)$ of  $CP^{3}$ by the following formula
\begin{equation}
c^1c^4=c^2c^3.
\end{equation}
In terms of the intrinsic coordinates $\zeta^1,\zeta^2,\zeta^3$,
in the chart with $c^1\neq 0$, i.e. $\xi^1=1$, the equation (11)
is
\begin{equation}
\zeta^1\zeta^2=\zeta^3.
\end{equation}


\subsection{Complex structure on ${\textbf C}P^n$}

Consider a complex manifold ${\cal M}$ with complex dimension
$\dim_{C}{\cal M}=n$ (in particular ${\textbf C}P^n$ ). We can
look at ${\cal M}$ as a real manifold with $\dim_{R}{\cal M}=2n$.
The real coordinates $(x^1,\dots x^{2n})$ are related to the
holomorphic $(\zeta^1,\dots \zeta^n)$ and anti-holomorphic
$(\bar\zeta^1,\dots \bar\zeta^n)$ coordinates via the following
formulas:
\begin{eqnarray}
(x^{\nu}+\imath x^{\nu+n})/\sqrt{2}&=&\zeta^{\nu},\qquad
\nu=1,2,\dots n, \nonumber\\(x^{\nu}-\imath
x^{\nu+n})/\sqrt{2}&=&\bar\zeta^{\nu},\qquad \nu=1,2,\dots n,
\end{eqnarray}
and
\begin{eqnarray}
 q^{\nu}\equiv x^{\nu}&=&(\zeta^{\nu}+\bar\zeta^{\nu})/\sqrt{2},\qquad \nu=1,2,\dots
 n,\nonumber\\
p^{\nu}\equiv
x^{\nu+n}&=&(\bar\zeta^{\nu}-\bar\zeta^{\nu})/\sqrt{2},\qquad
\nu=1,2,\dots n.
\end{eqnarray}

 The tangent space $T_{x}{\cal M}$ is spanned by $2n$ vectors:
\begin{equation}
\{{\partial\over \partial q^1},\dots {\partial\over \partial
q^n},{\partial\over \partial p^1},\dots {\partial\over \partial
p^n}\}
\end{equation}
or by the basis
\begin{equation}
\{{\partial\over \partial \zeta^1},\dots {\partial\over \partial
\zeta_n},{\partial\over \partial \bar \zeta^1},\dots
{\partial\over
\partial \bar \zeta_n}\}.
\end{equation}

An almost complex structure on a real $2n$-dimensional manifold is
given by a $(1,1)$ tensor ${\textbf J}$ satisfying ${\textbf
J}^2={\textbf 1}$, i.e.  $J_{c}^{a} J_{b}^{c}=-\delta_b^a$.
Locally, the almost complex structure $J$ is given in the real
coordinates by the following matrix
\begin{equation}
\pmatrix{0&-{\bf 1}\cr
 {\bf 1}&0},
\end{equation}
where ${\bf 1}$ is $n$-dimensional unit matrix.
 If the real $2n$ manifold is
actually a complex manifold, like in our case, the almost complex
structure is defined globally and is called the complex structure.


\subsection{Riemannian structure on ${\textbf C}P^n$}

Hermitian scalar product induces a complex Euclidean  metric on
${\textbf C}^N$. The metric induced on ${\textbf C}P^n$ is the
Fubini-Study metric, and is given, in
 $(\zeta,\bar\zeta)$ coordinates, using an
 $n\times n$ matrix with following entries
 \begin{equation}
 g_{\mu,\bar\nu}(\zeta,\bar\zeta)={\delta_{\mu,\nu}(1+\zeta\bar\zeta)-\zeta^{\mu}\bar\zeta^{\nu}\over
 (1+\zeta\bar\zeta)^2},\quad \mu,\nu=1,2\dots n,
 \end{equation}
where $\zeta\bar\zeta\equiv \sum_{\mu}^n
\zeta^{\mu}\bar\zeta^{\mu}$.

The Fubini-Study metric in
 $(\zeta,\bar\zeta)$ coordinates is then given by $2n\times 2n$
 matrix
\begin{equation}
{\cal G}(\zeta,\bar \zeta)={1\over 2}\pmatrix{0&g_{\mu,\bar\nu}\cr
g_{\bar\mu,\nu} &0}.
\end{equation}
In the real coordinates the Fubini-Study metric is given by the
standard transformation formulas
\begin{equation}
{\cal G}_{i,j}(q,\bar p)={\cal G}_{k,l}(\zeta(q, p),\bar\zeta(q,
p)){\partial Z_i\over \partial X_k}{\partial Z_j\over \partial
X_l},
\end{equation}
where we used $Z=(\zeta^1,\dots \bar\zeta_n$ and $X=(q^1\dots
p_n)$.

In the example of two qubits the Fubini-Study metric on ${\textbf
C}P^3$ is $$2{\cal G}=$$
\begin{equation}
\pmatrix{0&0&0&{(1+\zeta\bar\zeta)-\zeta^{1}\bar\zeta^{1}\over
 (1+\zeta\bar\zeta)^2}&{-\zeta^{1}\bar\zeta^{2}\over
 (1+\zeta\bar\zeta)^2}&{-\zeta^{1}\bar\zeta^{3}\over
 (1+\zeta\bar\zeta)^2}\cr
0&0&0&{-\zeta^{2}\bar\zeta^{1}\over
 (1+\zeta\bar\zeta)^2}&{(1+\zeta\bar\zeta)-\zeta^{2}\bar\zeta^{2}\over
 (1+\zeta\bar\zeta)^2}&{-\zeta^{2}\bar\zeta^{3}\over
 (1+\zeta\bar\zeta)^2}\cr
0&0&0&{-\zeta^{3}\bar\zeta^{1}\over
 (1+\zeta\bar\zeta)^2}&{-\zeta^{3}\bar\zeta^{2}\over
 (1+\zeta\bar\zeta)^2}&{(1+\zeta\bar\zeta)-\zeta^{3}\bar\zeta^{3}\over
 (1+\zeta\bar\zeta)^2}\cr
{(1+\zeta\bar\zeta)-\zeta^{1}\bar\zeta^{1}\over
 (1+\zeta\bar\zeta)^2}&{-\zeta^{2}\bar\zeta^{1}\over
 (1+\zeta\bar\zeta)^2}&{-\zeta^{3}\bar\zeta^{1}\over
 (1+\zeta\bar\zeta)^2}&0&0&0\cr
{-\zeta^{1}\bar\zeta^{2}\over
 (1+\zeta\bar\zeta)^2}&{(1+\zeta\bar\zeta)-\zeta^{2}\bar\zeta^{2}\over
 (1+\zeta\bar\zeta)^2}&{-\zeta^{3}\bar\zeta^{2}\over
 (1+\zeta\bar\zeta)^2}&0&0&0\cr
{-\zeta^{1}\bar\zeta^{3}\over
 (1+\zeta\bar\zeta)^2}&{-\zeta^{2}\bar\zeta^{3}\over
 (1+\zeta\bar\zeta)^2}&{(1+\zeta\bar\zeta)-\zeta^{3}\bar\zeta^{3}\over
 (1+\zeta\bar\zeta)^2}&0&0&0},
\end{equation}
Transformation to the real coordinates, by application of the
formula (20), gives
\begin{equation}
\pmatrix{{b\over a} &-{p^1p^2+q^1q^2\over a} &-{p^1p^3+q^1q^3\over
a }&0&{p^1q^2-p^2q^1\over a}&{p^1q^3-p^3q^1\over a} \cr
-{p^1p^2+q^1q^2\over a }&{b\over a} &{p^2p^3+q^2q^3\over a
}&{p^2q^1-p^1q^2\over a}&{0}&{p^2q^3-p^3q^2\over a}\cr
-{p^1p^2+q^1q^2\over a }&{p^2p^3+q^2q^3\over a }&{b\over a}
&{p^3q^1-p^1q^3\over a}&{p^3q^2-p^2q^3\over a} &0\cr
{0}&{p^2q^1-p^1q^2\over a}&{p^3q^1-p^1q^3\over a}&{b\over
a}&-{p^1p^2+q^1q^2\over a}&-{p^1p^3q^1q^3\over a}\cr
{p^1q^2-p^2q^1\over a}&0&{p^3q^2-p^2q^3\over
a}&-{p^1p^2+q^1q^2\over a}&{b\over a}&-{p^2p^3+q^2q^3\over a}\cr
{p^1q^3-p^3q^1\over a}&{p^2q^3-q^2p^3\over a}
&0&-{p^1p^3-q^1q^3\over a}&-{p^2p^3+q^2q^3\over a}&{b\over a}},
\end{equation}
where
 $$
a=(p^1)^2+(p^2)^2+(p^3)^2+(q^1)^2+(q^2)^2+(q^3)^2+2,\quad
b=(p^1)^2+(p^3)^2+(q^1)^2+(q^3)^2+2.
 $$
 Obviously, ${\cal G}$ is positive definite and symmetric.

\subsection{Symplectic structure on ${\textbf C}P^n$}

The Hermitian scalar product on ${\textbf C}^N$ is also used to
define the symplectic structure on ${\textbf C}^N$ and  this
induces  the symplectic structure on ${\textbf C}P^n$. The
symplectic structure is the closed nondegenerate two form $\Omega$
on ${\textbf C}P^n$, which is, in $(\zeta,\bar\zeta)$ coordinates
given by
\begin{equation}
\omega=\imath g(\zeta,\bar\zeta)_{\mu,\bar\nu}
d\zeta^{\mu}\wedge\bar\zeta^{\nu}
\end{equation}
where $g_{\mu,\bar\nu}$ is the Fubini-Study metric (18). In real
coordinates, the symplectic structure is given by $
{\Omega}(q,p)=J{\cal G}(q,p)$ where ${\cal G}(q,p)$ is given by
(20) and $J$ by (17).

The symplectic form on the two qubits state space is in the real
bases given by the product of matrices (17) and (22). The results
is
$${\Omega}=$$
\begin{equation}
\pmatrix{0&{-p^2q^1+p^1q^2\over a^2}&{-p^3q^1+p^1q^3\over
a^2}&-{b\over a^2}&{p^1p^2+q^1q^2\over a^2}&{p^1p^3+q^1q^3\over
a^2}\cr
 {p^2q^1-p^1q^2\over a^2}&0&{p^2q^3-p^3q^2\over a^2}&{p^2p^1+q^1q^2\over a^2}&-{b\over a^2}&{p^2p^3+q^2q^3\over a^2}\cr
  {p^3q^1-p^1q^3\over a^2}&{p^3q^2-p^2q^3\over a^2}&0&{p^1p^3+q^1q^3\over a^2}&{p^2p^3+q^2q^3\over a^2}&-{b\over a^2}\cr
{b\over a^2}&-{p^2p^1+q^1q^2\over a^2}&-{p^1p^3+q^1q^3\over
a^2}&0&{p^1q^2-p^2q^1\over a^2}&{p^1q^3-p^3q^1\over a^2}\cr
-{p^1p^2+q^1q^2\over a^2}&{b\over a^2}&-{p^2p^3+q^2q^3\over
a^2}&{p^2q^1-p^1q^2\over a^2}&0&{p^2q^3-p^3q^2\over a^2}\cr
-{p^1p^3+q^1q^3\over a^2}&-{p^2p^3+q^2q^3\over a^2}&{b\over a^2}
&{p^3q^1-p^1q^3\over a^2}&{p^3q^2-p^2q^3\over a^2}&0}
\end{equation}

\section{Quantum Hamiltonian dynamical system on $ {\textbf{C}}P^n$ }

The Schroedinger equation on $C^N$ is in some basis $\{|\psi_i>,\>
i=1,2 \dots N\}$ given by:
\begin{equation}
\imath {dc^i\over dt}=<\psi_j| H|\psi_i>c^j.
\end{equation}
In the real coordinates this equation assumes the form of a
Hamiltonian dynamical system on $R^{2N}$ with a global gauge
symmetry corresponding to the invariance $|\psi>\rightarrow \exp
(i x) |\psi>$. Reduction with respect to this symmetry results in
the Hamiltonian system on $ {\textbf{C}}P^n$, considered as a real
manifold with the symplectic structure given by (23). The Hamilton
equation on $ {\textbf{C}}P^n$, that are equivalent to the
Schroedinger equation (25), are
\begin{equation}
 {dx^l\over dt}= 2\Omega^{l,k}\nabla_k H(x),
\end{equation}
where $\Omega^{l,k}$ is the inverse of the symplectic form, and
$H(x)$ is given by the normalized quantum expectation of the
Hamilton's operator $<\psi |H|\psi>/<\psi|\psi>$ expressed in
terms of the real coordinates (14). For example, the hamiltonian
(7) is given in terms of the real coordinates $q^i\equiv x^i,\>
p^i\equiv x^{i+n},\> i=1,\dots n$ by
\begin{equation}
H={\omega\over a}[2-(p^3)^2-(q^3)^2]+{\mu\over a}
(p^1p^2+q^1q^2+\sqrt{2}q^3).
\end{equation}
and  the symmetric hamiltonian (9) is given by
\begin{equation}
H={\omega\over a} [2-(p^3)^2-(q^3)^2]-{\mu\over a}
[(p^1)^2+(p^2)^2+(q^1)^2+(q^2)^2-(p^3)^2-(q^3)^2-2]
\end{equation}

 The Hamilton's
equations (26) with the hamiltonian (27) and the symplectic form
(24) assume the following form
\begin{eqnarray}
\dot q^1&=&-2\omega p^1+\mu p^2-\mu(p^3q^1+p^1q^3)/\sqrt{2}\nonumber\\
\dot q^2&=&-2\omega p^2+\mu p^2-\mu(p^3q^2+p^2q^3)/\sqrt{2}\nonumber\\
\dot q^3&=&-4\omega p^3-\sqrt{2}\mu p^3q^3\nonumber\\
\dot p^1&=&2\omega q^1-\mu q^2+\mu(q^3q^1-p^1p^3)/\sqrt{2}\nonumber\\
\dot p^2&=&2\omega q^2-\mu q^1+\mu(q^3q^2-p^2p^3)/\sqrt{2}\nonumber\\
\dot p^3&=&4\omega q^3+\mu ((q^3)^2-(p^3)^2-2)/\sqrt{2}.
\end{eqnarray}

 The equations of motion with the symmetric hamiltonian (28) on ${\textbf C}P^3$ are
 quite simple
\begin{eqnarray}
\dot
q^1&=&-2(\omega+\mu)p^1\nonumber\\
\dot
q^2&=&-2(\omega+\mu)p^2\nonumber\\
\dot
q^2&=&-4\omega p^3\nonumber\\
\dot
p^1&=&2(\omega+\mu)q^1\nonumber\\
\dot
p^2&=&2(\omega+\mu)q^1\nonumber\\
\dot p^3&=&4\omega q^3.
\end{eqnarray}

\subsection{ Quantum Hamiltonian system with imposed separability constraints}

Dynamics of a constrained Hamiltonian system is usually described
by the method of Lagrange multipliers \cite{Dirac},\cite{Arnold}.
Consider a Hamiltonian system given by a symplectic manifold
${\cal M}$ with the symplectic form $\Omega$ and the Hamilton's
function $H$ on ${\cal M}$. Suppose that besides the forces
described by $H$ the dynamics of the system is affected also by
forces whose sole effect is to constrain the motion on a
submanifold ${\cal N}\in {\cal M}$ determined by a set functional
relations
\begin{equation}
f_1(q,p)=\dots f_k(q,p)=0
\end{equation}
The method of Lagrange multiplies assumes that the dynamics on
${\cal N}$ is determined by the following set of differential
equations
\begin{equation}
\dot X=\Omega(\nabla X,\nabla H'), \qquad H'=H+\sum_j^k,
\lambda_jf_j
\end{equation}
which should be solved together with the equations of the
constraints (31). The Lagrange multipliers $\lambda_j$ are
functions of $(p,q)$ that are to be determined from the following,
so called compatibility, conditions.
\begin{equation}
\dot f_l=\Omega(\nabla f_l,\nabla H')
\end{equation}
on ${\cal N}$. The equations (33) uniquely determine the functions
$\lambda_1(p,q),\dots \lambda_k(p,q)$ if and only if the matrix of
Poison brackets $\{f_i,f_j\}=\Omega(\nabla f_i,\nabla f_j)$ is
nonsingular. If this is the case then all constraints (31) are
called primary, and ${\cal N}$ is symplectic manifold with the
symplectic structure determined by the so called Dirac-Poison
brackets
\begin{equation}
\{F_1,F_2\}'=\{F_1,F_2\}+\sum_{i,j}^k
\{f_i,F_1\}\{f_i,f_j\}^{-1}\{f_j,F_2\}
\end{equation}
 As we shall see, this is the case in the examples of pairs of
interacting qubits constrained on the manifold of separable states
that we shall analyze. On the other hand, if some of the
compatibility equations do not contain multipliers, than for that
constrain $\dot f_j=\{f_j,H\}=0$, which represents an additional
constraint. These are called secondary constraints, and they must
be added to the system of original constraints (31). If this
enlarged set of constraints is functionally independent one can
repeat the procedure. At the end one either obtains a
contradiction, in which case the original problem has no solution,
or one obtains appropriate multipliers $\lambda_k$ such that the
system (33) is compatible. In the later case the solution for
$\lambda_k$ might not be unique in which case the orbits of (32)
and (31) are not uniquely determined by the initial conditions.

Let us apply the formalism of Lagrange multipliers on the system
of two interacting qubits additionally constrained to remain on
the manifold of separable pure state.
 The real and imaginary parts of (12) give the two constraints in
 terms of real coordinates $(q^1,q^2,q^3,p^1,p^2,p^3)$
 \begin{equation}
 f_1=p^1p^2-q^1q^2+\sqrt {2} q^3, \qquad f_2=\sqrt {2}
 q^3-p^2q^1-p^1q^2
 \end{equation}
 The compatibility conditions (33) assume the following form
\begin{eqnarray}
 \dot f_1& =& \Omega(\nabla f_1,\nabla H)+\lambda_2 \Omega(\nabla f_1,\nabla f_2 )=0,\nonumber\\
 \dot f_2& =& \Omega(\nabla f_2,\nabla H)+\lambda_1 \Omega(\nabla f_2,\nabla
 f_1)=0.
 \end{eqnarray}
 where $\Omega$ is the symplectic form (24) and $\Omega(\nabla f_1,\nabla
 H)=\Omega^{a,b}\nabla_a f_1\nabla_b
 H$.

 The matrix of Poisson brackets $\{f_i,f_j\}$ on ${\cal N}$
is
\begin{equation}
 \pmatrix{ 0  & [2+(p^1)^2+(q^1)^2][2+(p^2)^2+(q^2)^2]/8\cr
-[2+(p^1)^2+(q^1)^2][2+(p^2)^2+(q^2)^2]/8 & 0}
 \end{equation}
 and is nonsingular. Thus the compatibility conditions can be solved for the Lagrange multipliers
 $\lambda_1(q,p),\lambda_2(q,p)$,
\begin{eqnarray}
\lambda_1&=&4\mu{4p^1p^2q^1q^2+[(q^1)^2-2][2+(p^2)^2-(q^2)^2]+(p^1)^2[(q^2)^2-(p^2)^2-2]\over
[2+(p^1)^2+(q^1)^2)^2(2+(p^2)^2+(q^2)^2]^2}\nonumber\\
\lambda_2&=&8\mu{(p^1)^2p^2q^2-p^2q^2[(q^1)^2-2]+p^1q^1[2+(p^2)^2-(q^2)^2]\over
[2+(p^1)^2+(q^1)^2)^2(2+(p^2)^2+(q^2)^2]^2}.\nonumber\\
 \end{eqnarray}
  Finally, the dynamics of the constrained system
is described by the equations (32) and (31) with
$\lambda_1(q,p),\lambda_2(q,p)$ and $f_1(q,p), f_2(q,p)$ given by
(38) and (35). For the Hamiltonian (27) the resulting equations of
motion for $q^1,q^2,p^1,p^2$ are
\begin{eqnarray}
 \dot q^1&=&-{4\mu p^1q^1q^2+2\omega p^1[2+(p^2)^2+(q^2)^2]\over 2+ (p^2)^2+(q^2)^2},\nonumber\\
  \dot q^2&=&-{4\mu p^2q^1q^2-2\omega p^2[2+(p^1)^2+(q^1)^2]\over 2+ (p^1)^2+(q^1)^2},\nonumber\\
   \dot p^1&=&{2\mu q^2 [(q^1)^2-(p^1)^2-2]+2\omega q^1[2+(p^2)^2+(q^2)^2]\over 2+ (p^2)^2+(q^2)^2},\nonumber\\
   \dot p^2&=&{2\mu q^1 [(q^2)^2-(p^2)^2-2]+2\omega q^2[2+(p^1)^2+(q^1)^2]\over 2+
(p^1)^2+(q^1)^2}.
 \end{eqnarray}
 The same procedure for the symmetric hamiltonian (28) results with
 the following equations of motion
\begin{eqnarray}
\dot q^1&=&{2\mu p^1[(p^2)^2+(q^2)^2-2)]-2\omega p^1[2+(p^2)^2+(q^2)^2]\over 2+ (p^2)^2+(q^2)^2},\nonumber\\
  \dot q^2&=&{2\mu p^2[(p^1)^2+(q^1)^2-2]-2\omega p^2[(2+(p^1)^2+(q^1)^2]\over 2+ (p^1)^2+(q^1)^2},\nonumber\\
   \dot p^1&=&{-2\mu q^1 [(q^2)^2+(p^2)^2-2]+2\omega q^1[2+(p^2)^2+(q^2)^2]\over 2+ (p^2)^2+(q^2)^2},\nonumber\\
   \dot p^2&=&{-2\mu q^2 [(q^1)^2+(p^1)^2-2]+2\omega q^2[2+(p^1)^2+(q^1)^2]\over 2+ (p^1)^2+(q^1)^2}.
 \end{eqnarray}
 There are also the equations expressing $\dot q^3$ and $\dot p^3$
 in terms of $q^1,q^2,p^1,p^2$, but the solutions of these are
  already given by the constraints.

\subsection{ Qualitative properties of the constrained dynamics of
two interacting qubits }

In this section we present the results of numerical analyzes of
the qualitative properties of the dynamics generated by the
constrained equations (40) and (39), corresponding to the quantum
Hamiltonians (28) with the $SO(2)$ symmetry  and (27)  without
 such symmetry.

 It is well known that any quantum system is integrable
  when considered as the Hamiltonian dynamical system on the
  symplectic space ${\cal H}$, and that the reduction on the symplectic manifold $P{\cal
  H}$ preserves this property. This is simply a consequence of the form of the
 quantum  Hamiloton's function, which is always defined as the mean value of the Hamiltonian operator.
   Contrary to the case of classical
  Hamiltonian systems, the symmetry of the physical system has no
  relevance for the property of integrability in the Hamiltonian
  formulation of the Schroedinger equation. We illustrate this fact, in
  figures 1a,b,  by projections on $(q^1,p^1)$ plane of a typical orbit for
  the symmetric and nonsymmetric hamiltonians of the pair of
  qubits. The motion on ${\textbf{C}}P^3$ in the symmetric case has further degeneracy
  compared with the nonsymmetric case, but both cases generate
  integrable, regular Hamiltonian dynamics.

\begin{figure}
  \includegraphics[height=.3 \textheight, width=0.5 \textheight]{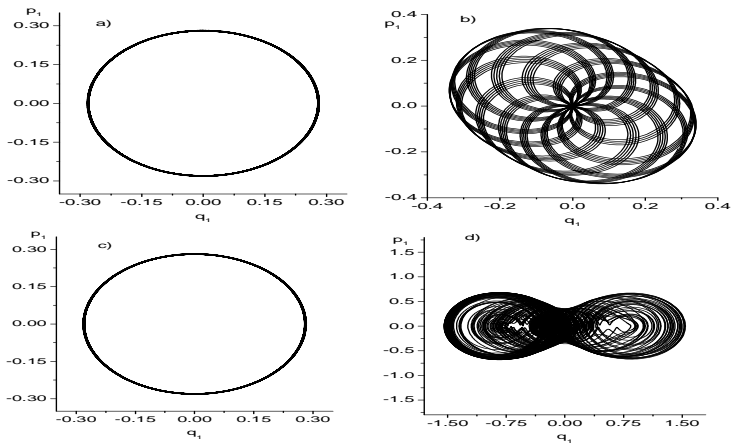}
  \caption{Projections on $(q_1,p_1)$ plane of a typical orbit
for the hamiltonian systems (28) (a) and (27) (b) on $
{\textbf{C}}P^3$ and on the submanifold of separable states (c)
for (40) and d for (39). The values of the parameters are
$\omega=1$ and $\mu=1.7$  }
\end{figure}

 On the other hand, the qualitative properties of the dynamics
 constrained by the separability conditions, are quite different.
 Typical orbits in the symmetric and nonsymmetric cases are
 illustrated in figure 1c,d. Symmetric dynamics  constrained by
 separability is still regular, while the
 nonsymmetric Hamiltonian generates the constrained dynamics with  typical chaotic
 orbits. This is further illustrated in figures 2, where
we show Poincar\' e surfaces of section, defined by $q^2=0, p^2>0$
and $H(p^1,q^1,p^2,q^2)=h$ for different values of the coupling
$\mu$. Obviously, the constrained system displays the transition
from predominantly regular to predominantly chaotic dynamics, with
all the intricate structure of the phase portrait, characteristic
for typical Hamiltonian dynamical systems. Thus, we can conclude
that the quantum system constrained on the manifold of separable
state behaves as typical classical Hamiltonian systems. If there
is enough symmetry, i.e. enough integrals of motion, the
constrained dynamics is integrable, otherwise the constrained
quantum dynamics is that of typical chaotic Hamiltonian system.

\begin{figure}
  \includegraphics[height=.3 \textheight, width=0.6 \textheight]{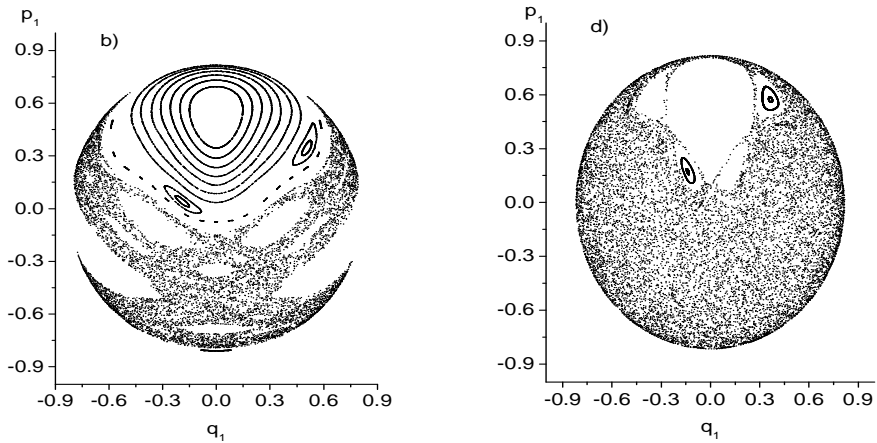}
  \caption{ Poincar\' e sections for the separability
 constrained non-symmetric quantum dynamics (39). The parameters are
$\omega=1,h=1.5$ and (a) $\mu=1.3$, (b) $\mu=1.7$}
\end{figure}

\section{ Summary and discussion}

We have studied Hamiltonian formulation of quantum dynamics of two
interacting qubits.
 Hamiltonian dynamical system on the state space ${\textbf{C}}P^3$ as the phase space, is integrable
 irrespective of the different symmetries of the quantum system. We have then studied
 the dynamics of the quantum Hamiltonian system constrained on the manifold of separable states.
 The main result of this analyzes, and of the paper, is that the quantum Hamiltonian system
 without symmetry generates nonintegrable chaotic dynamics on the set of separable states,
 while the  constrained symmetric dynamics gives an integrable system.
It is important to bare on mind that neither the system nor the
separable states that lie on an orbit of
 the constrained system
 have an underlining classical mechanical model. Thus,
 forcing a non-degenerate quantum system to remain on the manifold of
 separable states is enough to generate a dynamical system with typical properties of Hamiltonian chaos.

 Our analyzes of the separability constrained quantum dynamics has
 been rather formal. In order to inquire into possible
 interpretation of our results we need a model of a physical
 realization of the separability constraints. To this end we
 consider an open quantum system of two interacting qubits, whose
 dynamics satisfies the Markov assumption \cite{Brauer}, and we choose a
 Hermitian Lindblad operator of the following form
 \begin{eqnarray}
 L&=&l_{11} \sigma_+^1\sigma_-^1\otimes \sigma_+^2\sigma_-^2+
 l_{12} \sigma_+^1\sigma_-^1\otimes \sigma_-^2\sigma_+^2+
 l_{21} \sigma_-^1\sigma_+^1\otimes \sigma_+^2\sigma_-^2+
 l_{22} \sigma_-^1\sigma_+^1\otimes
 \sigma_-^2\sigma_+^2\nonumber\\
&=&\sum_{i,j=1}^2 l_{i,j} |i><j|^1\otimes |i><j|^2
 \end{eqnarray}
where $|1>\equiv |\uparrow>$ and $|2>\equiv |\downarrow>$.

The dynamics of a pure state of the open system under the action
of a Hamiltonian $H$ and the Linblad $\gamma L$ is described by
the following stochastic nonlinear  Schroedinger equation
\cite{Brauer},\cite{QSD}
\begin{eqnarray}
|d\psi>&=&-i H |\psi>dt
+{\gamma^2\over 4}( L-<\psi|L|\psi>)^2|\psi>dt\nonumber\\
&+& \gamma ( L-<\psi|L|\psi>)|\psi> dW,
\end{eqnarray}
where  $dW$ is the increment  of complex Wiener c-number process
$W(t)$.

The equation (42) represent a diffusion process on a complex
Hilbert space, and is central in the "Quantum State Diffusion"
(QSD) theory of open quantum systems \cite{QSD}. It has been used
to study the systems of interacting qubits in various environments
for example in \cite{ja1},\cite{ja2}, and the effect of the
Linblad operator (41) on the entanglement between two qubits was
considered in \cite{Izrael}. The influence  of the non-Hamiltonian
terms of drift (proportional to $\gamma^2)$ and the diffusion
(proportional to $\gamma)$), with the Linblad operator of the form
(41), is to drive an entangled state towards one of the separable
states with the corresponding probability. This process occurs on
the time scale proportional to $\gamma^{-1}$. So, for large
$\gamma$ there occurs an almost instantaneous collapse of an
entangled state into a separable one. We believe that with a
proper choice of the parameters $l_{i,j}$ the long term dynamics
of a pure state described by (42) can have the same qualitative
properties as the separability constrained quantum dynamics. In
particular, the difference between the qualitative properties of
symmetric and nonsymmetric systems, reflected in the constrained
Hamiltonian system, should also manifest in the dynamics of (42)
for a proper choice of $l_{i,j}$. This expectations are supported
by figures 3, which illustrate the dynamics of
$(<\sigma_x^1>,<\sigma_y^2>)$ for the Hamiltonian operators (5)
and (8) as calculated using the constrained Hamiltonian equations
(39) and (40) (figures 3b and 3a ), or the QSD equation (42)
(figures 3d and 3c) for a particular choice of $l_{i,j}$ and large
$\gamma=5$. Of course, the choice of optimal values for $l_{i,j}$
should be according to some criterion, which is the problem we are
currently investigating.

\begin{figure}
  \includegraphics[height=.3 \textheight, width=0.5 \textheight]{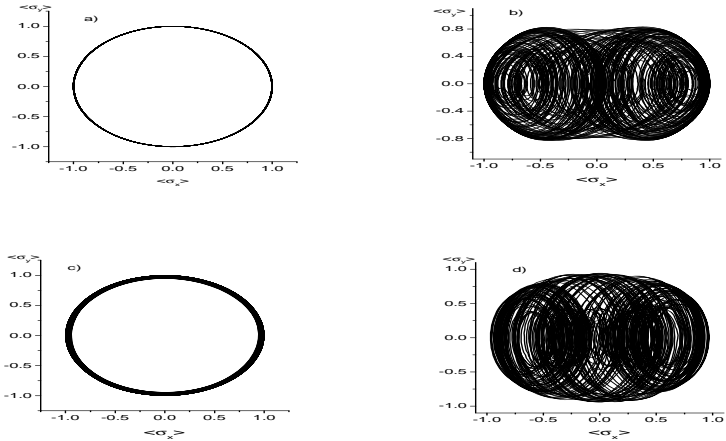}
  \caption{ Figures illustrate the dynamics of
$(<\sigma_x>,<\sigma_y>)$ for the constrained Hamiltonian systems
(40) (a) and (39) (b) and for the stochastic Schroedinger equation
(42) with the Linblad (41) and the hamiltonians (8) (c) and (5)
(d). the parameters are $\omega=1,\mu=1.7, \gamma=5$ and
$l_{1,1}=0.21,l{1,2}=0.21,l_{2,1}=0.215, l_{2,2}=0.205$. }
\end{figure}

The pair of coupled qubits, analyzed in this paper, is the
simplest quantum system exhibiting dynamical entanglement. We
intend to  investigate the effects of suppression of the dynamical
entanglement in systems with spacial degrees of freedom, obtained
by quantization of classically chaotic systems, for example a pair
of coupled nonlinear oscillators. In this case, the Hamiltonian
formulation of the quantum dynamics requires an
infinite-dimensional phase space, and the analyzes of the
separability constrained dynamics is more complicated. However, it
wold be interesting to compare the dynamics obtained by
separability constraints with that of some more standard
semi-classical approximation.

 {\bf Acknowledgements} This work is
partly supported by the Serbian Ministry of Science contract No.
141003. I should also like to acknowledge the support and
hospitality of the Abdus Salam ICTP.

\newpage
{\bf FIGURE CAPTIONS}

{\bf Figure 1 } Projections on $(q_1,p_1)$ plane of a typical
orbit for the hamiltonian systems (28) (a) and (27) (b) on $
{\textbf{C}}P^3$ and on the submanifold of separable states (c)
for (40) and d for (39). The values of the parameters are
$\omega=1$ and $\mu=1.7$

{\bf Figure 2 }  Poincar\' e sections for the separability
 constrained non-symmetric quantum dynamics (39). The parameters are
$\omega=1,h=1.5$ and (a)$\mu=1.1$, (b) $\mu=1.3$, (c) $\mu=1.5$
and (d) $\mu=1.7$

{\bf Figure 3 } Figures illustrate the dynamics of
$(<\sigma_x>,<\sigma_y>)$ for the constrained Hamiltonian systems
(40) (a) and (39) (b) and for the stochastic Schroedinger equation
(42) with the Linblad (41) and the hamiltonians (8) (c) and (5)
(d). the parameters are $\omega=1,\mu=1.7, \gamma=5$ and
$l_{1,1}=0.21,l{1,2}=0.21,l_{2,1}=0.215, l_{2,2}=0.205$.

\end{document}